\documentclass{ws-mpla}

\newcommand{\be}{\begin{equation}}
\newcommand{\ee}{\end{equation}}
\newcommand{\bea}{\begin{eqnarray}}
\newcommand{\eea}{\end{eqnarray}}

\newcommand{\bay}{\begin{array}}
\newcommand{\eay}{\end{array}}

\newcommand{\gev}{\ensuremath \! {\rm GeV}}

\newcommand{\tev}{{\rm TeV}}
\newcommand{\Mmes}{M_{\rm mess}}
\newcommand{\Nmes}{N_{\rm mess}}
\newcommand{\dilog}{\ensuremath{\mathrm{Li}_2}}

\begin{document}

\markboth{A. E. Blechman}
{RGM and the MRSSM}

\catchline{}{}{}{}{}

\title{R-SYMMETRIC GAUGE MEDIATION AND THE\\ 
MINIMAL R-SYMMETRIC SUPERSYMMETRIC STANDARD MODEL}

\author{\footnotesize ANDREW E. BLECHMAN}

\address{Department of Physics, University of Toronto, 60 St George Street\\
Toronto, Ontario M5S 1A7,
Canada\\
blechman@physics.utoronto.ca}

\maketitle

\pub{Received (Day Month Year)}{Revised (Day Month Year)}

\begin{abstract}
This is an invited summary of a seminar talk given at various institutions in the United States and Canada.  After a brief introduction, a review of the minimal $R$-symmetric supersymmetric standard model is given, and the benefits to the flavor sector are discussed.  $R$-symmetric gauge mediation is an attempt to realize this model using metastable supersymmetry breaking techniques.  Sample low energy spectra are presented and tuning is discussed.  Various other phenomenological results are summarized.

\keywords{Supersymmetric Effective Theories, Supersymmetric Phenomenology, Supersymmetry Breaking }
\end{abstract}

\ccode{PACS Nos.: 11.30.Hv, 11.30.Pb, 12.60.Jv}

\section{Introduction}	

Supersymmetry (SUSY) is a symmetry of spacetime that pairs each Standard Model (SM) Weyl fermion with a spin zero boson, and each boson with a Weyl fermion.  It is a very popular candidate for physics beyond the SM for a variety of reasons: it is known to solve the gauge hierarchy problem; it improves the agreement with Grand Unification; it can provide several potential candidates for dark matter; and it has several well studied theoretical properties, such as being the only extension of the Poincare algebra allowed by the Coleman-Mandula theorem, that make it a very attractive extension of physics.

However, as we do not see supersymmetry in nature, it must be that if weak scale SUSY exists it is spontaneously broken, thus giving masses to the supersymmetric partners.  It is well known that this will not do damage to the hierarchy problem's solution so long as the SUSY breaking operators are ``soft," meaning that they have a scaling dimension less than four.  Naively doubling the SM particle content, adding a second Higgs doublet, and writing down all the allowed soft SUSY breaking operators is called the Minimal Supersymmetric Standard Model (MSSM).

There are four types of allowed soft SUSY breaking operators that one can write down in the MSSM: hermitian scalar masses; holomorphic scalar masses; holomorphic trilinear terms (commonly called ``$A$ terms"); and Majorana gaugino masses.  There is only one holomorphic scalar mass term (involving the Higgs doublets) that is a singlet under the gauge symmetry, and this is called the $B$ term.  Putting it all together, there are 124 parameters in the MSSM.  Furthermore, most of these operators explicitly break the flavor symmetry of the SM, leading to large contributions to flavor changing neutral currents (FCNC) and CP violation that are in conflict with experiment.  The usual solution to this problem is to arbitrarily turn these couplings off, or insisting that they are small.  This is sometimes called the SUSY flavor puzzle.

An $R$-symmetry is a transformation that rotates fields within the same SUSY multiplet differently.  An example of this is the standard $R$-parity, where SM particles are even and SUSY particles are odd; this symmetry removes dangerous baryon and lepton violating operators from the MSSM.  Recently, it was realized in Ref.~\refcite{Kribs} that by introducing an additional $R$-symmetry beyond $R$-parity, there can be sizable flavor-violating operators while still not generating large FCNC or CP violation, so long as the gluinos are heavy.  In this review I will discuss this result, and then present a model called $R$-Symmetric Gauge Mediation (RGM), proposed in Ref.~\refcite{Amigo}, that seeks to realize this condition of an unbroken $R$-symmetry.  In everything that follows, a $U(1)_R$ symmetry is employed, although discrete symmetry groups work as well, such as $\mathbb{Z}_6$; however the usual $R$-parity of the MSSM is not strong enough.  Examples of this are discussed in Ref.~\refcite{Kribs}.

\section{Flavor Physics with an $R$-Symmetry}

Imposing a $U(1)_R$ symmetry on the MSSM, where gauginos and squarks have $R$ charges $r=+1$, and the Higgs scalar has $r=0$, there are three immediate consequences for our soft terms:

\begin{itemlist}
\item Majorana masses for the gauginos are now forbidden, but there can still be Dirac masses if one introduces new adjoint chiral multiplets.

\item $A$ terms are forbidden for the scalars.  This means that there are no $L-R$ mixing terms for the squarks or sleptons.

\item A $\mu$ term is forbidden for the Higgs sector, but a $B$ term can still be included.
\end{itemlist}
The absence of a $\mu$ term is a problem, and it means that the minimal Higgs sector is quite a bit more complicated in these models.  We will discuss this below.  This general class of models goes under the name of the Minimal $R$-symmetric Supersymmetric Standard Model (MRSSM).

\subsection{$\Delta F=2$ processes}

It was shown in Ref.~\refcite{Kribs} that the above consequences did much to relieve the flavor physics tension in the low energy spectrum.  Due to its tight experimental bounds, $K-\bar{K}$ mixing is the strongest constraint when studying the effects of SUSY flavor violation.  There are several diagrams that contribute to this mixing at one loop, but the most important numerically are those due to the exchange of gluinos.  Due to the absence of a Majorana mass for the gluino and $A$ terms for the squarks, several contributions that appear in the MSSM calculation cannot be generated at this order in perturbation theory.

It is assumed that the squarks are nearly degenerate in mass (this is a reasonable assumption for the first two generations in most SUSY breaking mediation scenarios), and we parametrize the off-diagonal elements of the mass matrix in terms of dimensionless numbers:\footnote{These quantities are usually referred to as $\delta_{LL},\delta_{RR}$ in the literature, but since there is no flavor mixing of the ``left-right" form, we drop the second $L(R)$ for notational simplicity.}
\be
\delta_L\equiv \frac{m_{\tilde{Q}12}^2}{M_{\tilde{q}}^2}\qquad
\delta_R\equiv \frac{m_{\tilde{d}12}^2}{M_{\tilde{q}}^2}~,
\label{qmix}
\ee
where $\tilde{Q}$ is the left-handed squark and $\tilde{d}$ is the right-handed down-type squark and $M_{\tilde{q}}$ is the universal squark mass.  These are the only relevant squark-mixing angles for $K-\bar{K}$ mixing.  The low energy effective Lagrangian has the form
\be
\mathcal{L}_{\rm eff}=\frac{\alpha_s^2(M_{\tilde{g}})}{216}\left(\frac{M_{\tilde{q}}^2}{M_{\tilde{g}}^4}\right)\sum_{n}C_n(\mu)\mathcal{O}_n(\mu)~, \label{Leff}
\ee
where the sum is over only the operators $\mathcal{O}_{1,4,5}$ and $\tilde{\mathcal{O}}_1$ as defined in Ref.~\refcite{Bagger}.  The absence of $\mathcal{O}_{2,3}$ are again due to the absence of $A$ terms.  In addition, this Lagrangian is suppressed by an additional factor of $(M_{\tilde{q}}/M_{\tilde{g}})^2$ over the MSSM calculation of Ref.~\refcite{Bagger}.  This is due to the Dirac nature of the gauginos, and suggests that the gluino should be heavy compared to the squarks in order to avoid flavor constraints.  The dependence of this effective theory on $\delta_{L,R}$ is in the coefficient functions $C_n$; the explicit form of these functions, including the leading log QCD corrections, is given in Ref.~\refcite{Blechman}.

\begin{figure}[tb]
\centerline{\scalebox{0.5}{
\includegraphics{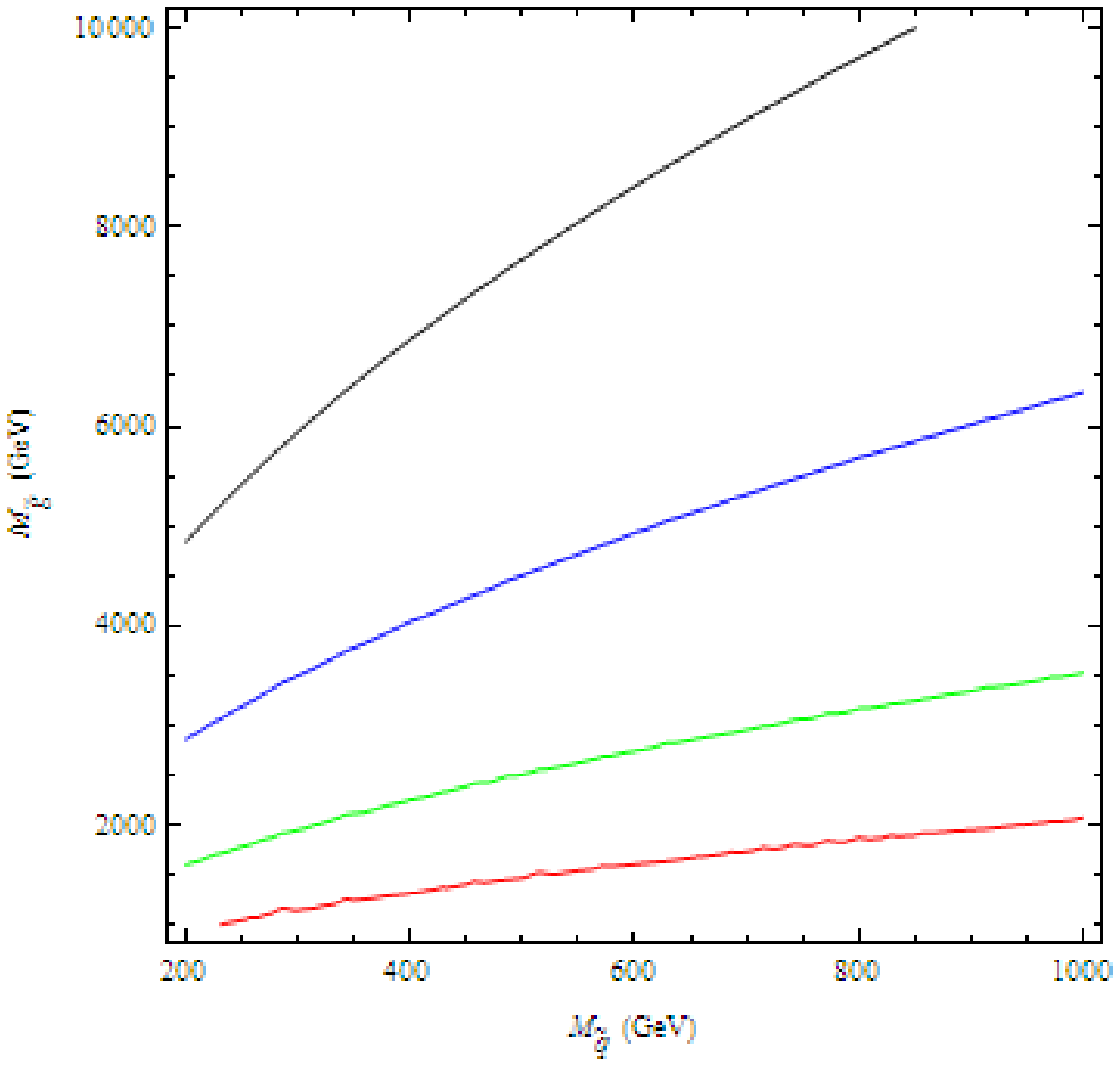}\hspace{0.15cm}
\includegraphics{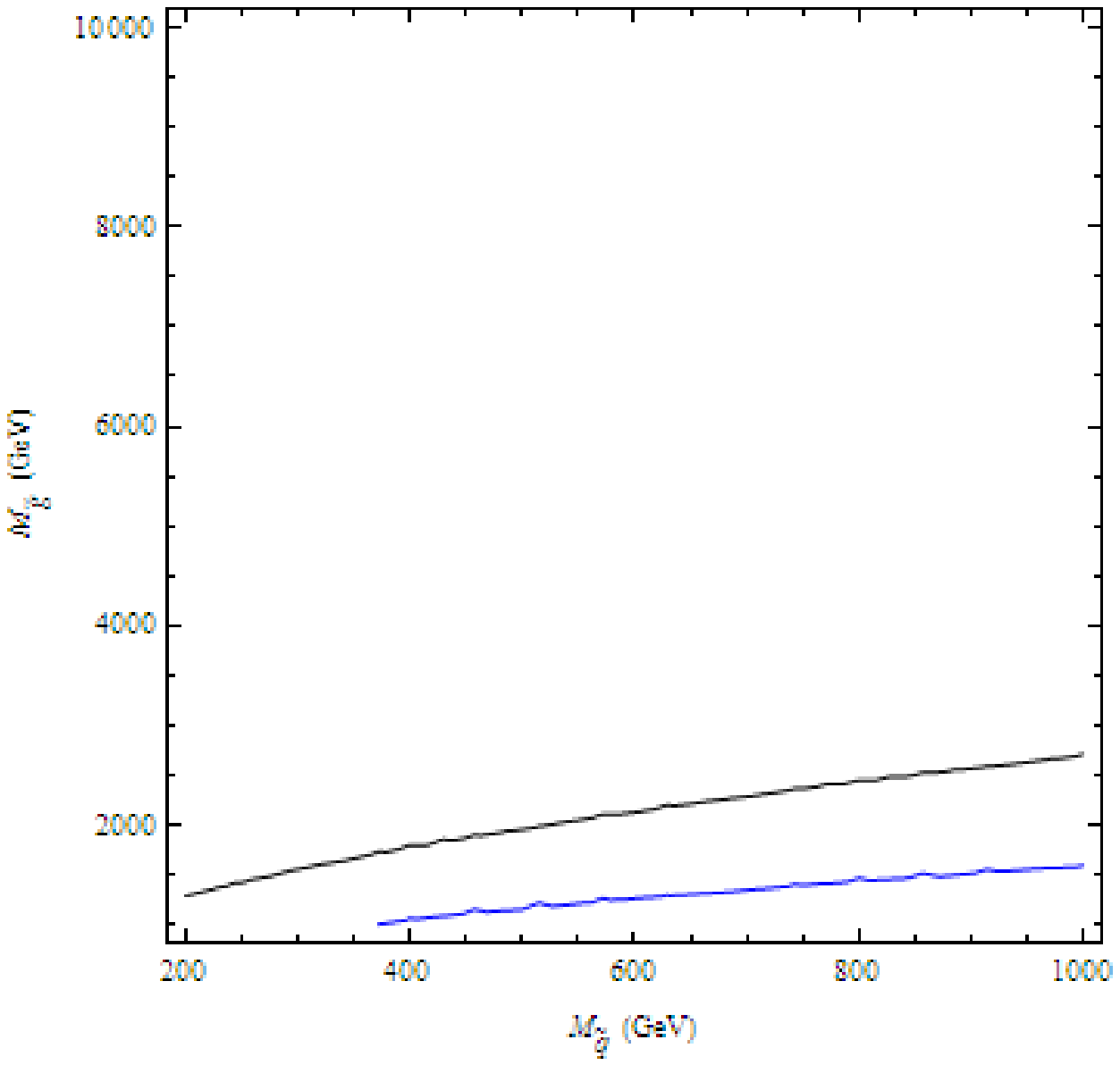}}}
\vspace*{8pt}
\caption{Exclusion plots in the $M_{\tilde{g}}-M_{\tilde{q}}$ plane, including QCD corrections.  The region below the lines is excluded.  For the plot on the left, from the bottom curve up: $\delta_L=\delta_R=0.01,0.03,0.1,0.3$.  The plot on the right has $\delta_L=0$ and $\delta_R=0.1,0.3$. \protect\label{results-equal}}
\end{figure}

From the plots in Figure \ref{results-equal} it can be seen that the QCD corrections do have a sizable effect on the results.  But even including these effects, $\delta_{L,R}$ can be quite large, even as large as 10\%, while still not violating the flavor constraints for LHC-accessible squark masses, so long as $m_{\tilde{g}}/m_{\tilde{q}}>5$.

We can also consider the effects on $\epsilon_K$ and CP violation.  Unfortunately, the QCD corrections are large for this observable, and force a constraint on $\arg(\delta_L\delta_R^*)< 10^{-3}$ for a typical point.  This suggests that some other mechanism must be at work to explain the lack of additional CP violation in this scenario.

One can repeat this analysis for other $\Delta F=2$ processes such as $B-\bar{B}$ mixing, but one finds much weaker constraints.  For example, in the $B$ system, we only require $m_{\tilde{g}}/m_{\tilde{q}}>2$ for $\delta=0.1$ to be within experimental bounds.  More details are shown in Ref.~\refcite{Kribs}.

\subsection{Other studies of flavor violation in the MRSSM.}

In this section, other flavor changing processes that were considered in Ref.~\refcite{Kribs} are briefly reviewed.  For more details, see their paper.

\begin{itemlist}
\item $\mu\rightarrow e\gamma$: In the MSSM, this process is dominated by diagrams involving $\mu$ and Majorana gaugino mass insertions.  Neither is present in the MRSSM, which causes many contributions to vanish at one loop.  For $\delta=0.1$, it is found that $m_{\tilde{B}}/m_{\tilde{l}}>2$ is sufficient to maintain experimental bounds.

\item $\varepsilon^\prime/\varepsilon$: The constraint can be written as
\be
{\rm Im}(\delta)<1.2\left(\frac{m_{\tilde{g}}}{500~\gev}\right)\left(\frac{x}{25}\right)~,\nonumber
\ee
where $x=(m_{\tilde{g}}/m_{\tilde{q}})^2$.  From Figure \ref{results-equal}, $\delta=0.1$ is consistent with $x>25$ so this is no constraint at all.

\item $B\rightarrow\tau\nu,~\mu^+\mu^-$:  In the MSSM, these processes can occur by radiatively generated couplings of the up-type Higgs to the down-type quarks.  However, these couplings are forbidden by PQ symmetry, which is a good symmetry in the MRSSM Higgs sector up to the $B$ term.  At one loop, the $B$ term does not contribute to these operators, and so these processes are automatically suppressed.

\item {\it Neutron EDM}: These are generated in the MRSSM through loops that involve the adjoint scalars that must come with the Dirac gauginos.  These scalars can get a holomorphic mass which we call $B_\phi$.\footnote{In Ref.~\refcite{Kribs}, they use the notation $M_{\tilde{g}}\equiv\sqrt{B_\phi}$.}  The calculated EDM for the neutron is
\be
|d_n|=(4\times 10^{-26}~e\cdot{\rm cm})\left(\frac{{\rm Im}\left(m_{\tilde{g}}\sqrt{B_\phi^{*}}\right)}{m_{\tilde{g}}^2}\right)\left(\frac{1~\tev}{m_{\tilde{g}}}\right)^2~. \nonumber
\ee
Experimental bounds give $|d_n|<6\times 10^{-26}~e\cdot{\rm cm}$ so there are no constraints here.

\item {\it Strong-CP}: Strong CP is a problem for the gauge sector.  Requiring the strong CP phase to obey $\bar{\theta}<10^{-9}$ requires $\arg\left(m_{\tilde{g}}\sqrt{B_\phi^{*}}\right)<10^{-7}$.  There is no clear way to accomplish this.

The good news is that unlike in the MSSM, the squark sector does not contribute to this phase, since squark contributions at one loop require the presence of $A$ terms and Majorana gaugino mass insertions, neither of which exist in the MRSSM.

\end{itemlist}

To conclude, the MRSSM does very well to explain the lack of FCNC, even though there could be sizable scalar mixing.  CP violation is a little harder to explain, but there is some evidence, such as $\varepsilon^\prime/\varepsilon$ and the lack of squark contributions to $\bar{\theta}$, that the MRSSM could help to alleviate the experimental constraints.

\section{$R$-Symmetric Gauge Mediation}

At first glance, the above model does have an apparent drawback: anytime you spontaneously break SUSY, you expect to also break any $R$-symmetry, at least in four dimensions.  The reason for this is due to the presence of a goldstino and a cosmological constant that necessarily comes in when SUSY is broken.  When you turn on supergravity, this goldstino is ``eaten" by the gravitino, which gets a mass by the SUSY version of the Higgs mechanism.  This mass is $R$-symmetry violating in $\mathcal{N}=1$ SUSY models.  Furthermore, this mass term will always feed into the low energy spectrum via anomaly mediation, inducing Majorana masses and $A$ terms, and so it looks like this model can never be realized in nature.

The solution is to employ a version of gauge mediation, which typically has a very light gravitino, typically order keV mass, which would render these troublesome contributions irrelevant.  However, ordinary gauge mediation breaks the $R$-symmetry as well, as it needs to do if it is to generate Majorana masses for the gauginos.  Recently there have been models proposed by Intrilligator, Seiberg and Shih (ISS) that have a metastable vacuum that spontaneously breaks SUSY but preserves an $R$-symmetry; see Ref.~\refcite{ISS}.  This vacuum can be made very stable, so it might provide a candidate for our MRSSM generation.

\subsection{Sample messenger sectors}

Previously, much effort has gone into finding ways to break the $R$-symmetry in ISS models so that the gauginos can get a mass and the model can be used to realize the MSSM; see, for example, Refs.~\refcite{ref1,ref2,ref3,ref4,ref5,ref6,ref7,ref8,ref9,ref10,ref11,ref12,ref13,ref14,ref15,ref16,ref17,Csaki}.  We are particularly interested in the minimal model that does this, discussed in Ref.~\refcite{Csaki}, which describes a SUSY-QCD model with $N_F=6,~N_C=5$, with a dual magnetic theory that is confining and the dual squarks get vacuum expectation values (vevs) that break the flavor symmetry to $SU(5)$; the $SU(6)$ symmetry is also explicitly broken to avoid Goldstone modes.  When this is done, there is a $SU(5)$ adjoint field $M$ ($r=+2$), two sets of $\mathbf{5}+\bar{\mathbf{5}}$ fields denoted $(N,\bar{N})$ and $(\varphi,\bar{\varphi})$ ($r=+2,0$ respectively), and several singlets.  We then gauge a $SU(3)\times SU(2)\times U(1)$ subgroup of this flavor symmetry and identify it with the SM gauge group.  The details of how this is done are given in Ref.~\refcite{Amigo}.  In addition to this, two additional adjoint superfields (called $\Phi$ and $M^\prime$, both $R$-neutral) are introduced, where $\Phi$ is only an adjoint under the SM gauge group.  These superfields will be used to give Dirac masses to the gauginos.  The $M^\prime$ superfield is needed to give mass to the otherwise massless fermions in the messenger sector meson field $M$.

In addition to the SUSY QCD operators in the magnetic ISS superpotential, it is also necessary to include couplings of the $\Phi$ field to the messengers that are allowed by symmetries:
\be
\label{yterms}
W_1= y\left( \bar\varphi \Phi N - \, \bar{N} \Phi \varphi\right)~.
\ee
These terms will be necessary to generate gaugino masses, and introduce a new parameter $y$.

The $N-\varphi$ fermions marry and aquire a Dirac mass which we call $M_{\rm mess}$.
The $(N,\bar{N})$ scalars get a SUSY preserving mass squared $M_{\rm mess}^2$, while the $(\varphi,\bar{\varphi})$ scalars develop a SUSY violating mass squared $(1\pm z)M_{\rm mess}^2$, where $0<z<1$ is a number that can be expressed in terms of the superpotential couplings of the ISS model.  

Due to the additional messengers with SM gauge quantum numbers, as well as the added adjoint superfields $\Phi$, the SM gauge groups develop Landau poles.  The presence of the extra adjoint $M^\prime$ is a real problem, because it brings down the QCD Landau pole ($\Lambda_3$) to near the messenger scale $M_{\rm mess}$.  Therefore, a generalization of this model was proposed that made no mention of the $SU(5)$ adjoint fields or singlets, but only the two sets of $\mathbf{5}+\bar{\mathbf{5}}$ fields and a SUSY breaking spurion.  A further generalization of these models is to allow for $N_{\rm mess}$ of these fields.  The new feature of this model over the usual gauge mediation is that  you need {\it two} sets of vectorlike fields rather than just one, in order to properly give Dirac masses to the gauginos.

In addition, there is a charge conjugation symmetry in the messenger sector, which exchanges barred and unbarred fields, and the gauge fields, gauginos and $\Phi$ superfield are odd under this parity.  This symmetry is important, since it forbids dangerous tadpole terms for the hypercharge adjoint, which is a singlet and could spoil the hierarchy as explained in Ref.~\refcite{BPR}.  This also forbids dangerous kinetic mixing of the SUSY breaking spurion with the hypercharge D term, which if allowed could generate tachyonic sleptons.  This symmetry also explains the choice of sign in Eq.~(\ref{yterms}).  It is analogous to the messenger parity that was proposed in Ref.~\refcite{Meade}.

\subsection{Low energy spectrum}

After integrating out the messengers, there are three parameters $(z,~y,~M_{\rm mess})$ for the ISS model, and also $N_{\rm mess}$ in the generalized model.  At this point, the model is akin to gauge mediated models, and in particular has no flavor puzzle, since gauge interactions are flavor universal and cannot generate flavor-violating operators.  However, as mentioned above, there is a Landau pole close to the messenger scale, and this means that our theory has a relatively low cutoff.  From the effective field theory point of view, this means that we should write down all operators consistent with the gauge and $R$- symmetries, including flavor-violating operators, suppressed by a cutoff $\Lambda\sim\Lambda_3/4\pi$.  Thus, we can insert flavor violation just above the messenger scale and ask whether the resulting low energy spectrum is consistent with the MRSSM flavor analysis summarized earlier.

These UV contributions to the SUSY breaking terms are all proportional to the scale
\be
M_{UV}=\frac{zM_{\rm mess}^2}{\Lambda}~.
\ee
The value of $\Lambda_3$ can be computed and gives a rough size for the soft mass operators up to coefficients $c^{ij}$ that are naively expected to be order unity.  The UV contributions to the real scalar masses are given by
\be
m_0^{ij}=c^{ij}M_{UV}~.\label{uvmass}
\ee
The UV operators that generate gaugino masses are suppressed, and there are also UV masses for the adjoint scalars $\Phi,~M,~M^\prime$, as explained in Ref.~\refcite{Amigo}.  It is the soft sfermion masses that we are most concerned with here.

In addition, there are gauge mediated contributions.  The soft masses get the usual two loop gauge mediated flavor diagonal contributions, but only from the $(\varphi,\bar{\varphi})$ sector, since the $(N,\bar{N})$ sector does not break SUSY.  So even though it looks like there are generally $2N_{\rm mess}$ messengers, there are actually only $N_{\rm mess}$ as far as the soft scalar loops are concerned.  The RGM mass squared for the sfermions from the gauge group $a=1,2,3$ is given by
\be
m_0^{2}=2C^{(a)}_F\left(\frac{\alpha_a}{4\pi}\right)^2M_{\rm mess}^2F(z)~,
\label{IRscalarmass}
\ee
where $C^{(a)}_F = (N^2-1)/2N$ for $SU(N)$ and $\frac{3}{5}Y^2$ for $U(1)_Y$ and:
\be
F(z)=(1+z)\left[\log(1+z)-2\dilog \left(\frac{z}{1+z}\right)+\frac{1}{2}\dilog\left(\frac{2z}{1+z}\right)\right]+(z\rightarrow -z)~.
\ee

\begin{figure}[tb] 
    \centering
    \includegraphics[width=4in]{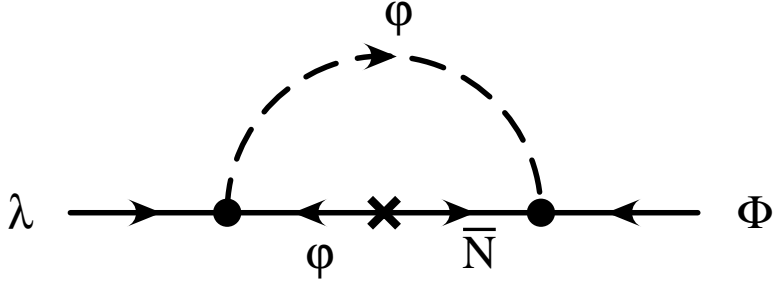} 
    \caption{One of the diagrams contributing to the 1-loop gaugino mass.  The other graphs are obtained by different choices of $\varphi$, $\bar{\varphi}$ $N$, and $\bar{N}$ running in the loop.}
    \label{fig:gauginomass}
 \end{figure}

The gaugino mass is quite different.  It is given by the loop in Figure \ref{fig:gauginomass}, with the result
\be
m_{1/2}=\frac{gy}{16\pi^2}M_{\rm mess}R(z)~, 
\label{IRgauginomass}
\ee
where:
\be
R(z)=\frac{1}{z}\left[(1+z)\log (1+z)-(1-z)\log (1-z)-2z\right]~,\label{gauginomassfun}
\ee
Note the dependence on $gy$ rather than the usual $g^2$ of ordinary gauge mediation.  There is also a similar diagram that generates masses for the adjoint scalars proportional to $y^2$.  We will say more about this below.

At this point, we come across the first problem of trying to realize the MRSSM flavor scenario.  Looking at Figure \ref{results-equal}, the gaugino had to be roughly five times heavier than the squarks.  This ratio is
\be
\frac{m_{1/2}}{m_0^{IR}} =\frac{1}{\sqrt{2C_F}}\left( \frac{y}{g}\right)\left(\frac{R(z)}{\sqrt{F(z)}}\right)~.
\label{IRgauginoscalarratio}
\ee
The ratio $R/\sqrt{F}$ is strictly less than $1$: for $z=0.99$, $|R/\sqrt{F}| =.64$.  Thus, in order to solve the supersymmetry flavor puzzle using the MRSSM  within an ISS supersymmetry-breaking-cum-mediation sector, we must have a large  Yukawa coupling $y$.

One point that should be remembered here is that the large ratio of gaugino to squark mass should appear at the gluino scale, while the ratio in Equation (\ref{IRgauginoscalarratio}) is at the messenger scale.  This effect does help, since there is a window of supersoft running where the scalars do not get loop contributions from the gauginos as described in Ref.~\refcite{Fox}, while the gauginos themselves run heavy in the IR.  Unfortunately, this effect is naively too small to be of much help due to the heaviness of the adjoint scalars discussed below, but it should be included in more detailed studies of these models.

The tension is relieved somewhat by going to our generalized model of $N_{\rm mess}$ messengers, where the above fraction is enhanced by a factor of $\sqrt{N_{\rm mess}}$.  In addition, the generalized model removes the extra adjoint fields, and this raises $\Lambda_3$ and also helps ease the tension between UV and gauge mediated contributions to scalar masses.

Below are three sample spectra: an ISS model with relatively small Yukawa; an ISS model with a large Yukawa; and a case of the generalized model.  In all cases $z=0.99$ -- there is nothing very special about this point, but $z$ should be close to unity to maximize the SUSY breaking.   In all three tables, only the IR contributions to squark and slepton masses are shown.  The UV contributions can be estimated by knowing the corresponding Landau Poles, which are computed to be $\Lambda_3=(8,~10,~50)\times 10^3~\tev$.  Also included in these tables are the light messenger masses; see Ref.~\refcite{Amigo} for details on these particles.
\begin{table}[h]
\tbl{Spectrum for ISS-RGM with $y=2$.\label{tab:smallyuk}}
{\begin{tabular}{@{}c|cc||cc@{}} \toprule
$SU(3)$ & $m_{\tilde{q}}$ & $1400~\gev$ & $m_{\tilde{g}}$ & $880~\gev$\\ \hline\hline
$SU(2)$ & $m_{\tilde{l}}$ & $360~\gev$& $m_{\tilde{W}}$ & $520~\gev$\\ \hline\hline
$U(1)$ & $m_{\tilde{e^c}}$ & $160~\gev$ & $m_{\tilde{B}}$ & $370~\gev$\\ \hline\hline
Messenger& $M,M^\prime, \tilde{\Phi}$ & $15~\tev$ & $m_{-}$ &  $10~\tev$\\
sector & $\Mmes$ & $100~\tev$ & $m_\xi$ & $3100~\gev$
\end{tabular}
}
\end{table}
\begin{table}[h]
\tbl{Spectrum for ISS-RGM with $y=8$.\label{tab:largeyuk}}
{\begin{tabular}{@{}c|cc||cc@{}} \toprule
$SU(3)$ & $m_{\tilde{q}}$ & $1300~\gev$ & $m_{\tilde{g}}$ & $3500~\gev$\\ \hline\hline
$SU(2)$ & $m_{\tilde{l}}$ & $350~\gev$& $m_{\tilde{W}}$ & $2100~\gev$\\ \hline\hline
$U(1)$ & $m_{\tilde{e^c}}$ & $160~\gev$ & $m_{\tilde{B}}$ & $1500~\gev$\\ \hline\hline
Messenger& $M,M^\prime, \tilde{\Phi}$ & $13~\tev$ & $m_{-}$ &  $10~\tev$\\
sector & $\Mmes$ & $100~\tev$ & $m_\xi$ & $13~\tev$
\end{tabular}
}
\end{table}
\begin{table}[h]
\tbl{Spectrum in the generalized model for $y=3$ and $N_{\rm mess}=6$.\label{tab:general}}
{\begin{tabular}{@{}c|cc||cc@{}} \toprule
 $SU(3)$ & $m_{\tilde{q}}$ & $1900~\gev$ & $m_{\tilde{g}}$ & $5300~\gev$\\ \hline\hline
$SU(2)$ & $m_{\tilde{l}}$ & $620~\gev$& $m_{\tilde{W}}$ & $3500~\gev$\\ \hline\hline
$U(1)$ & $m_{\tilde{e^c}}$ & $290~\gev$ & $m_{\tilde{B}}$ & $2600~\gev$\\ \hline\hline
Messenger sector& $\Mmes$ & $80~\tev$ & & 
\end{tabular}
}
\end{table}

Notice that the large Yukawa ISS model, though it has a moderately reasonable spectrum for the MRSSM flavor solution, requires a $y$ so large that perturbativity is called into question.  This shows that an ISS-RGM scenario does not lend itself to the MRSSM, at least perturbatively, although the generalized model is much better.  We will see more evidence for this conclusion below.

\subsection{Estimates of fine-tuning}

This model has two forms of fine tuning:
\begin{itemlist}
\item To make the squarks light enough compared to the gaugino, there must be a UV-IR cancellation in the diagonal soft masses.

\item To satisfy flavor constraints, there must be a tuning of the off-diagonal mass terms in the UV contribution.
\end{itemlist}

If $m_0$ is the physical scalar mass, and $m_{IR}$ is the RGM contribution, then we can write
\be
c_D\sim\frac{m_0^2-m_{IR}^2}{M_{UV}^2}~,
\ee
where $c_D$ is the diagonal element of the UV mass matrix in Equation (\ref{uvmass}).  Looking at the tables in the previous section, we can use this formula and plug in values for $M_{UV}$ and typical values of a few hundred GeV for $m_0$ to show that $c_D\sim 10^{-2}$ for ISS models, and $c_D\sim 1$ for the generalized models.  Once again, it looks like ISS-RGM is not realizing the MRSSM, although it should be emphasized that this result should not be taken too seriously, since the large $y$ limit is not trustworthy.

The second form of tuning comes from the condition that we would like $c_D\sim c_{OD}$ to solve the flavor puzzle.  Using the notation above, and letting $\delta_L=\delta_R\equiv\delta$, we can estimate the size of $c_{OD}$
\be
c_{OD}=\delta\left(\frac{m_0}{M_{UV}}\right)^2~.
\ee
Putting the expressions for $c_D$ and $c_{OD}$ together gives us a good test for flavor tuning
\be
t\equiv\left|\frac{c_{OD}}{c_D}\right|=\frac{\delta}{|1-(m_{IR}/m_0)^2|}\label{tuningeq}~.
\ee
Notice that this is independent of $M_{UV}$.

This formula shows that flavor tuning is somewhat unavoidable, even in the generalized models.  Using the above results for $K-\bar{K}$ mixing, we can construct Table \ref{tab:tuning}.

\begin{table}[h]
\tbl{Size of the flavor tuning for the MRSSM spectra considered above.\label{tab:tuning}}
{\begin{tabular}{@{}c|cc||c@{}} \toprule
$ $ & $m_0$ & $\delta$ & $t$ \\
\hline
ISS with Large $y$ & 600 GeV & 0.05 & 1.4\%  \\
General Model & 1 TeV & 0.07 & 2.7\%
\end{tabular}
}
\end{table}

Again, ISS-RGM shows itself to be tuned, but it seems that even the generalized RGM model cannot avoid a tuning of a few percent, although it is still much better than the MSSM.  In addition, the authors of Ref.~\refcite{Luo} have studied the effects of adding a hidden sector $D$ term in addition to this mechanism and showed that this can also help to eliminate some of the flavor tuning.

\section{Phenomenology}

In this section we briefly review some of the phenomenology of RGM models potentially relevant for the LHC.

\subsection{Scalar adjoint physics}

The scalar adjoints in $\Phi$ get real and holomorphic masses both from the UV contributions as well as from RGM.  In this section we will only consider the RGM contributions.

The masses can be calculated from diagrams very similar to Figure \ref{fig:gauginomass}, and were calculated independently in Refs.~\refcite{Amigo,Goodsell} with the result
\bea
m_\phi^2&=&\frac{y^2}{16\pi^2}\Mmes^2 R_s(z)~, \label{madj}\\
B_\phi&=&\frac{y^2}{16\pi^2}\Mmes^2 R(z)~, \label{Badj}
\eea
where
\be
R_s(z)=\frac{1}{z}\left[(1+z)^2\log(1+z)-(1-z)^2\log(1-z)-2z\right]~,
\ee
and the $z$ dependence in (\ref{Badj}) is the same as in the gaugino mass.  These terms are the same order of magnitude, but it can be shown that for any value of $z$, $|B_\phi|<m_\phi^2$ and so there are no tachyons.  Furthermore, $B_\phi$ is negative, so the scalar adjoint will be lighter than the pseudoscalar adjoint.  This can be very interesting because it is the scalar adjoint that can be produced at the LHC through gluon fusion.  In principle, both can be pair produced as well, and this would be an excellent signal of these models.  This possibility was considered in Ref.~\refcite{Plehn}.

Unfortunately, these masses are one loop masses, not the usual two loops that is common for scalar masses in gauge mediation.  The masses are enhanced by roughly a factor of $\sqrt{4\pi/\alpha}$ over the gauginos, and so they would naively be too heavy to be seen at the LHC.  The way out of this is due to a cancellation in the scalar mass between the real and holomorphic terms.  There can also be potential cancellation between the UV and RGM contributions.  The extent to which this is reasonable and not fine tuned is currently being explored.

The other possibility is to consider the electroweak adjoints rather than the color adjoints.  These are harder to produce since they don't couple to gluons, but they might be produced via associated production, similar to the Higgs boson.  Furthermore, light electroweak gauginos are not ruled out by flavor constraints, unlike the gluino, and so the electroweak scalar adjoints might be light enough to produce.  This possibility is also being studied.

\subsection{Higgs physics}

The Higgs sector of the MRSSM is much more complicated than the MSSM.  It can be thought of as containing two hypermultiplets $(H_u,R_u)$ and $(H_d,R_d)$ in $\mathcal{N}=2$ SUSY
\be
\label{wr}
\delta W = \mu_u H_u R_u +  \lambda^u_1 H_u \Phi_1 R_u + \lambda^u_2 H_u \Phi_2 R_u + (u \rightarrow d)~.
\ee
The $H$ superfields have $R$ charge $r=0$ and the $R$ superfields have $r=+2$.  In particular, notice the new form of the $\mu$ term.  There are now two of them, $\mu_u$ and $\mu_d$.  Another amusing thing to notice about the $\mu$ terms is that they preserve a PQ symmetry.  In the MSSM, the PQ symmetry is broken by {\it both} the $\mu$ term and the $B$ term; now only the $B$ term breaks it.  This means that, unlike the case of the NMSSM, the $\mu$ and $B$ terms cannot be generated by the same physics!

Due to the supersoft nature of the model, the Higgs $D$ term vanishes in the limit that the soft mass terms for the adjoint scalars vanish, as explained in Ref.~\refcite{Fox}.  This makes electroweak symmetry breaking difficult to accomplish.  Including the adjoint scalars, the Higgs sector contains 24 fields.  Attempts to analyze the spectrum are underway and should be released soon.\cite{Amigo2}

\subsection{Neutralino/chargino physics}

In ordinary GM the gravitino is the LSP and it is so here as well, with a mass $m_{3/2}\sim 1$ keV.  In Ref.~\refcite{KMR}, the neutralino and chargino mass matrices are analyzed in various limits.  It was found that in many regions of the MRSSM parameter space, the charginos are the NLSP, with a chargino-neutralino mass splitting of as much as 30 GeV for large $\tan\beta$, as opposed to the MSSM which sees a much smaller mass splitting ($\sim 3~\gev$) at {\it low} $\tan\beta$.  Furthermore, all SUSY particles can decay through a cascade involving the charged wino, whose lower mass bound is set at 101 GeV by LEP-II.  This leads to new and interesting collider signals that might be of interest for the LHC; for example, if the gravitino is heavier than 100 eV then the charginos can escape the detector leaving charged tracks.  This would be a very distinctive signal.

\subsection{Top physics}

Single top production is a strong candidate for finding new flavor physics at the Tevatron and the LHC.  In Ref.~\refcite{KMRtop}, single top events have been studied as a signal for flavor violation in the squark mass matrix, both in the MSSM and in the MRSSM.  Specifically, they looked at stop pair production, where one stop decays to a top quark and a neutralino, and the other stop decays to a quark and a neutralino, where this second quark is not a top quark.  They found that flavor violation in this channel is maximized for a Bino LSP, and the best case for detection is for the neutralino to decay inside the detector to a photon and a graviton, leading to a nearly background-free two hard photon signal.  If the chargino is the NLSP as described above, then this signal will be suppressed, but in the regions of parameter space where this is not the case, this provides another powerful test of the MRSSM.  

The reader is referred to Refs.~\refcite{KMR,KMRtop} for more details on both of these collider possibilities.

\section{Conclusions}
By imposing a stronger $R$-symmetry than the $R$-parity of the MSSM, it was realized that you can have large off-diagonal squark and slepton mass terms while not violating any of the FCNC constraints of precision electroweak studies.  This opens the door to a variety of new flavor-violating signals at the weak scale that are usually assumed to be suppressed in ordinary SUSY phenomenology studies.  With the LHC about to turn on, this possibility is exciting.  It would imply a new spectrum of sparticles, new scalar fields with both strong and weak couplings to matter, and an entirely new Higgs sector.  The chargino and neutralino sectors also holds surprises, such as a light, long lived chargino that could leave a spectacular signal at the LHC.

RGM is one way to realize such scenarios.  The generalized model with $\Nmes$ messenger pairs generates a relatively untuned spectrum (although some tuning is required), and a Landau pole is generated in the right place for flavor-violating physics to enter the low-energy picture and provide an MRSSM solution to the flavor puzzle.


\section*{Acknowledgments}

I would like to thank all my collaborators for their contributions to this subject, and also Erich Poppitz and Santiago De Lope Amigo for suggestions on this manuscript.  I also thank Siew-Phang Ng for the plots in Figure \ref{results-equal}  I acknowledge support from the National Science and Engineering Research Council (NSERC) of Canada.

\end{document}